\documentclass[prb,showpacs,twocolumn,amsmath]{revtex4}
\usepackage{latexsym}

\usepackage{graphicx}% Include figure files

\newcommand{\be}{\begin{equation}}

\newcommand{\ee}{\end{equation}}

\newcommand{\reff}[1]{(\ref{#1})}

\newcommand{\giro}[1]{\stackrel{\circ}{#1}}

\begin{document}

\title{Negative ${\mathcal R}$-parity scissors modes in the  Two-Rotors Model }

\author{ Fabrizio Palumbo}

\affiliation{
INFN Laboratori Nazionali di Frascati, 00044 Frascati, Italy}

\begin{abstract}

In a previous work I constructed ${\mathcal R}$-negative scissors states in the  Two-Rotors Model  in which the rotors are two-sided classical  bodies. Such states have $|K|= 1,0$ components and  negative space parity. Here I extend this work including  reflections  in a plane through the rotors symmetry axes obtaining also ${\mathcal R}$-negative states of positive parity. I then associate reflections and   ${\mathcal R}$-operations with corresponding operations on internal noncollective variables of the rotors. I evaluate the em transition strengths of these states   and compare them  with the corresponding strengths of  ${\mathcal R}$-positive scissors modes.

\end{abstract}

\maketitle

{\bf Introduction}. The Two-Rotors Model  (TRM) describes the dynamics of two rigid bodies rotating with respect to each other under an attractive force around their centers  of mass. It was devised as a  model for deformed atomic nuclei, in which case  the rigid bodies represent the proton and neutron systems~[\onlinecite{LoIu}].  The collective $1^+$ state, the scissors mode, predicted by  this model was firstly observed~[\onlinecite{Bohle}]   in $^{154 }$Gd and then in all deformed atomic nuclei [\onlinecite{LoIu1}].  Recently also the prediction[\onlinecite{LoIu}] of the $J=2$ member of its rotational band has been confirmed [\onlinecite{Beck}].

By analogy similar collective excitations were predicted in several other systems~[\onlinecite{Guer}] and clearly observed in Bose-Einstein condensates~[\onlinecite{Mara}] and very recently in dipolar quantum droplets of Dysprosium atoms~[\onlinecite{Ferr}] and in supersolids ~[\onlinecite{Rocc,Tanz}]. Moreover an application of the TRM to the evaluation of the magnetic susceptibility of single domain magnetic nanoparticles stuck in rigid matrices has given results  compatible with a vast body of experimental data with an agreement  in some cases surprisingly good~[\onlinecite{Hata}].
Later   it was found that, according to the Brink hypothesis~[\onlinecite{Brin}], scissors modes live also on excited nuclear levels~[\onlinecite{Krti, Adek}].

There is by now an extensive literature on the subject~[\onlinecite{LoIu1}]: in all these papers the rotors were assumed to have axial symmetry and to be invariant  under inversion of their axes. Their wf should therefore be even or odd under such an inversion, but only even states were considered.  Inversion  is defined as a rotation-reflection~[\onlinecite{Hame}]: Rotation by $\pi$ about an axis perpendicular to the axis of the rotor assumed to have axial symmetry, called  ${\mathcal R}$, and reflection,  which I call $\sigma$,  in the plane perpendicular to the axis of this rotation. Inversion on a rotor  coincides with a parity operation; parity  on the two-rotors system is the product of inversions on the rotors. 

Recently  I investigated~[\onlinecite{Palu}] the effect of relaxing the restriction to positive ${\mathcal R}$ parity and I  constructed ${\mathcal R}$-negative  states of the TRM. Because I was actually interested in nuclear states of negative parity, I assumed the rotors  to be even under the reflection $\sigma$ that was therefore altogether omitted in the notation. The negative parity states so constructed  turned out to be  superpositions of $K= \pm1, 0$ components. 

Evidence of $K$ mixing in the rare earths was reported already long ago [\onlinecite{Zilg}], but my interest in the subject was renewed by a very recent work~[\onlinecite{Beck1}] in  which  a $1^-$ state in $^{164}$Dy at the energy of the scissors mode has been observed. In addition it was found that the states of a  $1^+$ doublet, also at the energy of the scissors mode,  have  $|K|=0, 1$ components of comparable amplitude. It was not reported, however,  whether the $1^-$ state has or not a $K=0$ component. So even though these findings may suggest a connection with my results just outlined, they  do not bear at present a strict connection with them. Nevertheless they open perhaps a more realistic perspective for my work on negative parity scissors modes, of which they suggest an extension.  In my  work indeed negative parity  states have vanishing electric dipole transitions, because roto-oscillations of classical objects cannot generate an electric dipole. But these classical objects are two-sided rotors with a quantal ${\mathcal R}$ number, which might require  a role of noncollective internal variables. 
The wave functions in collective models depend in fact on  collective as well as noncollective internal variables. The latter include the particle position variables  subject to constraints which  compensate for the redundancy.  Classical formulations are the method of canonical transformation~[\onlinecite{Vill}] and the method of redundant variables~[\onlinecite{Sche}]. In several cases it is acceptable to keep the single particle variables without any   constraint. For instance the noncollective internal variables appear in transition matrix elements and they were systematically set equal to 1 in all works on the TRM and in particular in my  paper~[\onlinecite{Palu}] on negative parity states, but this may be not  appropriate in the presence of negative ${\mathcal R}$ symmetry.  
The inversion of the axis of one rotor, in fact,   can be obtained acting either on the collective coordinates or performing a reflection on the internal noncollective  coordinates in the plane perpendicular to the symmetry axis of the rotor through the center,  and one must require that the two operations give the same result. For negative ${\mathcal R}$ symmetry the rotor is a two-sided object which may need a quantal description.

In the present work I evaluate the em transition strengths of scissors modes with {\it negative ${\mathcal R}$-symmetry and  both negative and positive parity},  allowing the rotors to be odd under the reflection $\sigma$ defined above.  I thus find that the negative parity scissors mode I previously constructed can decay to the ground state by an electric dipole transition. This is a purely quantal effect, because  roto-oscillations of classical rotors cannot produce any electric moment. In addition I evaluate magnetic dipole strengths of scissors modes of positive parity and negative ${\mathcal R}$ symmetry. 
 
I will distinguish the rotors by the labels $1,2$, that will be later identified with the proton and neutron fluids respectively.

{\bf The  Two-Rotors Model}. The classical  Hamiltonian of the TRM is
\be
H =\frac{1}{2{\mathcal J}_1}  {\vec L}_1^2 +  \frac{1}{2{\mathcal J}_2} {\vec L}_2^2 +V \label{classic}
\ee
where ${\vec L}_1, {\vec L}_2  $,  ${\mathcal J}_1, {\mathcal J}_2$  are the angular momenta and moments of inertia of  the rotors  and $V$  the  potential interaction, which depends only  on the angle  $2 \theta$ between the rotor axes. 
 If one denotes by  ${\hat \zeta}_1, {\hat \zeta}_2 $ the  unit vectors in the direction of the rotors axes, $ \cos(2\theta)= {\hat \zeta}_1 \cdot {\hat \zeta}_2$. It is convenient to replace them 
 by  the Euler angles $\alpha, \beta, \gamma$  that describe the orientation of the system of the rotors as a whole plus the variable $\theta$.  Euler angles are associated  with the direction cosines of the axes of the intrinsic  frame  
 \begin{eqnarray}
{\hat \xi} &=&\frac{{\hat \zeta}_2 \times {\hat \zeta}_1}{ 2 \sin\theta}, \, 
{\hat \eta}=\frac{{\hat \zeta}_2 - {\hat \zeta}_1}{2 \sin\theta}, \,
{\hat \zeta}=\frac{{\hat \zeta}_2 + {\hat \zeta}_1}{2 \cos \theta}\,.
\end{eqnarray}
 One  must then express all the operators  in terms of the new variables.  Setting
 \be
{\vec I}={\vec L}_1 + {\vec L}_2, \,\,\, 
{\vec S}= {\vec L}_1 - {\vec L}_2 
\ee
where ${\vec I}$ is the total orbital angular momentum acting on the Euler angles, while ${\vec S}$ is not an angular momentum,  one has the representation~[\onlinecite{LoIu}]
\be
S_{\xi}= i \frac{\partial}{\partial \theta} , \,\,\,  S_{\eta}= -\cot \theta I_{\zeta}, \,\,\,
 S_{\zeta}=-\tan \theta  I_{\eta} \,.
\ee 
%Notice that an incorrect definition of $S_1$ was given in Ref.[\onlinecite{DeFr}]; The derivation of the above will be reported shortly \cite{Palu0}.
In terms  of these variables  inversion of the rotor is defined as a {\it rotation-reflection[\onlinecite{Hame}]: rotation through $\pi$ about the $\xi$-axis followed by a reflection $\sigma$ in the   $\eta$-, $\zeta$-plane}
\be
{\mathcal I}_1=\sigma_1 \mathcal {R}_1 \,,  \,\,\, 
 {\mathcal I}_2= \sigma_2  \mathcal {R}_2 
 \label{inversion}
\ee
where
\be
\mathcal {R}_1 = R_{\zeta}(\pi) R_{\xi}(\frac{\pi}{2}) R_{\theta}\,,  \,\,\,\,
 \mathcal {R}_2 =    R_{\eta}(\pi) R_{\xi}(\frac{\pi}{2}) R_{\theta}
\ee
 and
\be
R_{\theta} f(\theta)= f(\pi/2 - \theta)= \giro{f}(\theta)\,.
\ee 
Physical states must be eigenstates of the  inversion operators ${\mathcal I}_1, {\mathcal I}_2$ 
\be
{\mathcal I}_1\Psi = r_1 \Psi\,, \,\,\,  {\mathcal I}_2\Psi= r_2 \Psi\,,  \,\,\, r_i=\pm1\,.
\ee
 For the whole system parity is the product of the inversions on the rotors
\be
{\mathcal P}= {\mathcal I}_1{\mathcal I}_2  \label{pariy}\,.
\ee

{\bf  The  intrinsic Hamiltonian of the Two-Rotors Model}. For the potential it was assumed a harmonic approximation, with restoring force constant  $C$. The range of $\theta$ was divided into 2 regions:  $ 0 <\theta < \frac{\pi}{4}, \, \frac{\pi}{4} < \theta < \frac{\pi}{2} $ and 2 functions were defined, $s_I, s_{II}$, which are respectively equal to 1 inside the 2 regions and zero outside.  Introducing the moment of inertia of the two-rotors system as a whole
\be
{\mathcal J}= {\mathcal J}_1 {\mathcal J}_2/ ({\mathcal J}_1 + {\mathcal J}_2). \label{J}
\ee
and the parameters
\be
\theta_0^2= \frac{1}{\sqrt{ {\mathcal J} C}},  \,\,\,\, 
 \omega = \sqrt \frac{C}{ {\mathcal  J }}    
\ee
the potential was written
\be
V  \approx {1\over 2}\omega \Big[ x^2 s_I + y^2 s_{II} \Big] \label{potential} 
\ee
where $ x=\theta/ \theta_0, \, y=(\pi / 2 -\theta) /  \theta_0$.  The harmonic approximation can be justified for  $\theta_0 << 1$. In the rare earth region $\theta_0 < 10^{-1}$.

The quantized  Hamiltonian,  setting $\hbar = c=1$, can then be put  in the approximate form~[\onlinecite{DeFr}] 
\be
H= \frac{{\vec I}^2 }{2{\mathcal J}} + {\mathcal H}_{0} +  {\mathcal H}_{1}
\ee
where
\begin{eqnarray}
 {\mathcal H}_{0}&=& \frac{1}{2} \omega \, s_I \left[- \frac{\partial^2}{\partial x^2} + \frac{1}{x^2} \Big(I_{\zeta}^2 - \frac{1}{4} \Big) + x^2 \right] 
\nonumber\\
&+& \frac{1}{2} \omega \, s_{II} \left[- \frac{\partial^2}{\partial y^2} + \frac{1}{y^2} \Big(I_{\eta}^2 - \frac{1}{4} \Big) + y^2 \right] 
\end{eqnarray}
\begin{eqnarray}
 {\mathcal H}_{1}&=&C_0 \, s_I \left[- \frac{1}{2 x}  \Big(I_{\zeta}I_{\eta} +I_{\eta} (I_{\zeta}\Big) +i I_{\xi}
 \frac{\partial}{\partial x} \right] 
\nonumber\\
&+& C_0\, s_{II} \left[ - \frac{1}{2 y}  \Big(I_{\zeta}I_{\eta} +I_{\eta} (I_{\zeta}\Big) - i I_{\xi}
 \frac{\partial}{\partial y} \right] 
\end{eqnarray}
and $C_0=   ({\mathcal J}_2 - {\mathcal J}_1)/ {{(\mathcal J}_1 + {\mathcal J}_2)} \,  \omega \theta_0$.  The difference in sign in front of the derivatives in ${\mathcal H}_{1} $ is due to the fact that $ \frac{\partial}{\partial \theta}  =(\theta_0)^{-1} (s_I\frac{\partial}{\partial x} - s_{II}\frac{\partial}{\partial y})$. $ {\mathcal H}_{1}$ is a small term that until now has been neglected, but its effect should be considered for  ${\mathcal R}$-negative states because they are opposite in sign depending on which rotor is ${\mathcal R}$-negative.
The total Hamiltonian  is the sum of the rotational Hamiltonian of the two-rotors system as a whole plus an intrinsic Hamiltonian. The above  expression clearly shows that {\it the roto-oscillations occur about both the $\zeta$- and $\eta$-axes}.
 
 The lowest eigenfunctions of ${\mathcal H}_{0} $, normalized to $1/2$, are
\begin{eqnarray}
\varphi_K(x) &=&  \frac{1}{\sqrt {(K+1)  \theta_0}} \, x^{K+ \frac{1}{2}} \, e^{-{ 1\over 2} x^2}
\nonumber\\
\varphi_K(y) &=&  \frac{1}{\sqrt {(K+1)  \theta_0}} \, y^{K+ \frac{1}{2}} \, e^{-{ 1\over 2} y^2}
\end{eqnarray}
in region I  and  II with energy eigenvalues $\omega(1+|K|)$ respectively. I will omit the functions $s_I, s_{II}$ when I will explicitly write the arguments of the functions. The wf of the ground  and the $2^+$ states in region I are  
\be
\mathcal {F}^0_{00}  \phi_0(x) \,,  \,\,\,\,\mathcal {F}^2_{M0}  \phi_0(x) 
\ee
where
\be
\mathcal {F}^I_{MK} = \sqrt{{2I+1}\over 16 (1+ \delta_{K0})\pi^2 } \left( {\mathcal D}^I_{MK} + ( -)^I  {\mathcal D}^I_{M-K}   \right)   \label{italic}\,.
\ee
The corresponding wf in region II involve the eigenstates of $  I_{\eta}^2  $ with zero eigenvalue which are
\be
 \mathcal {F}^0_{00}\,,  \,\,\,\,\, \mathcal {G}^2_{M0}=\frac{1}{2} ( \mathcal {F}^2_{M0} + \sqrt 3 \,  \mathcal {F}^2_{M2}) \,.  \label{normal}
\ee
 Neglecting $ {\mathcal H}_{1}$,  the condition of positive $ \mathcal {R}_1$ and $ \mathcal {R}_2$ symmetry gives the entanglement between the 2 regions
\begin{eqnarray}
\Psi_{0^+  } &=&\mathcal {F}^0_{00} (\phi_0(x) +  \phi_0(y) )
  \nonumber\\
\Psi_{2^+ M } &=&\mathcal {F}^2_{M0}  \phi_0(x) - \mathcal {G}^2_{M0}\phi_0(y).  \label{2^+}
\end{eqnarray}
The mixing  of terms with different K-values  is due to the breaking of axial symmetry of the nucleus because of the relative motion of the rotors axes.

In the present work I study  the em decay of ${\mathcal R}$-negative scissors modes of positive and negative parity to the above states. I will compare with the corresponding strengths of the ${\mathcal R}$-positive scissors modes whose wf, {\it neglecting }${\mathcal H}_1 $ are~[\onlinecite{LoIu}] 
\be
\Psi_{1^+ M }=\mathcal {F}^1_{M1} (\phi_1(x) - \phi_1(y))\,. \label{gs}
\ee

{\bf ${\mathcal R}$-negative  scissors modes}. The structure of ${\mathcal R}$-negative  scissors modes, always  neglecting $ {\mathcal H}_1$,  is [\onlinecite{Palu}] 
 \be
\Lambda^{\pi}_{M K} = F^1_{MK}(\alpha, \beta, \gamma) \chi_{ K  }(\theta) f^{\pi}   \label{lambda}
\ee
where
\be
F^1_{M 1}= \sqrt{{3}\over 16 \pi^2 } \left( {\mathcal D}^1_{M1} + {\mathcal D}^1_{M-1}   \right)\,,  \,\,\,\,\,
F^1_{M 0} = \mathcal {F}^1_{M0} \,,
 \ee
 \be
\chi_{1} = s_I  \, \varphi_1\,, \,\,\,\,\,\chi_{0 } =s_{II} \varphi_1 \,.
\ee
The factor  $f^{\pi} $  is a function of internal noncollective variables  which I introduce here according to the previous discussion. Notice the difference in sign wr to the italic $\mathcal {F}$'s of Eq.\reff{italic}.
There are 2 degenerate states distinguished by the $K$-quantum number: The states $\Lambda_{M 1 }$ and  $\Lambda_{M 0 }$ describe a precession  about the $\eta, \zeta$-axis and live in region $I, II$ respectively. They have the same energy as the ${\mathcal R}$-positive scissors modes, apart from the unknown energy necessary to excite one rotor to its ${\mathcal R}$-negative and $\sigma=\pm 1$ state, which will be neglected.

The action of a separate ${\mathcal R}$-operation  on the above wave functions is 
\begin{eqnarray}
{\mathcal R}_1 \Lambda_{1M 1 } =  i \, \Lambda_{1M 0 },\,\,\,\,\,
{\mathcal R}_1 \Lambda_{1M 0} = -\, i \Lambda_{1M 1 } &&
\nonumber\\
{\mathcal R}_2 \Lambda_{1M 1} = - i \Lambda_{1M 0} , \,\,\,
{\mathcal R}_2 \Lambda_{1M 0 }=  i  \, \Lambda_{1M 1 }. &&
\end{eqnarray}
One should notice that the operators  ${\mathcal R}_1, {\mathcal R}_2$ do not merely invert the axes $\zeta_1, \zeta_2$, but they also interchange the  $\Lambda$'s with each other. So the $\Lambda$'s are not eigenstates of 
${\mathcal R}_1, {\mathcal R}_2$ separately, but by repeated application of the above equations one finds that their total 
${\mathcal R}$ symmetry is negative
\be
{\mathcal R} \,  \Lambda_{1M 1 } ={\mathcal R} \, \Lambda_{1M 0 } =-1 \label{total}
\ee
where ${\mathcal R} = {\mathcal R}_1 {\mathcal R}_2$. 
If each rotor must be in a definite ${\mathcal R}$ state, however, so must be the total wf.  One can construct such a wf, normalized to 1,  in terms of the $\Lambda$'s
 \be
\Psi^{\pi} _{1 M  r_1} =  \left(  \Lambda_{M 1} + i r_1 \Lambda_{M 0 } \right)  f^{\pi}
\label{parity}
\ee
where 
\be
{\mathcal R}_1 \Psi_{1 M  \, r_1} =  r_1 \, \Psi_{1  M \,  r_1} \,, \,\,\,
{\mathcal R}_2 \Psi_{1  M  \,  r_1} = - r_1 \, \Psi_{1  M  \,  r_1}
 \label{negative}
\ee
that show that in these states the rotors must have opposite ${\mathcal R}$ symmetry. 
{\it In the above notation it is understood that all the scissors modes in this paper are $ \mathcal {R}$-negative, with the exception of Eq.\reff{positive}}. Following the notation of Ref.[\onlinecite{Beck1}] I will sometimes set
\be
\Psi^{+} _{1   r_1=-1} = 1^+_{sc,1}\,,   \,\,\,\, \Psi^{+} _{1  r_1=+1} = 1^+_{sc,2}\,.
\ee
Instead of the quadruplets [\onlinecite{Palu}] with fully classical rotors there is a doublet, a state in which rotor 1 is ${\mathcal R}$-even and rotor 2 is ${\mathcal R}$-odd and vice versa. For each such state there is a doublet whose members have opposite reflection symmetry $\sigma$ in the $\zeta$, $\eta$ plane and then opposite parity. (The actual existence of such states obviously depends  on the internal physics of the rotors, whether only one or both or none of them can be excited to a $\mathcal {R}$ and/or $\sigma$-negative  state).

In the  $1^+_{sc,1} $ it is rotor 1 which is ${\mathcal R}$-negative.  For consistency as said at the beginning  $f^{\pi}$ must then satisfy the condition $f^{\pi}(x_1,y_1,z_1)= - f^{\pi}(x_1,-y_1,-z_1)$, where $x_1, y_1z_1$ are the components of the position vector inside rotor 1. The parity of the total wf is then equal to the inversion symmetry of rotor 1, which is negative/positive if this rotor is even/odd  under the reflection $\sigma_1$.  Therefore I set
\be
f^{\pi}(\vec{r}_1) = s(z_1) g^{\pi} (x_1). \label{noncollective}
\ee
where $g^{\pi} (x_1)=1$ for $\pi=-1$ and $g^{\pi} (x_1)= s(x_1)$ for $\pi=1$. 
The wf of noncollective variables of  the ground and the $ 2^+$ states of Eqs.\reff{2^+} are set  equal to 1. The modifications for  $1^+_{sc,2}$ are obvious. 

The magnetic moments of rotor i  in the intrinsic frame are
\begin{eqnarray}
&&({\mathcal M}_i^{\pi})' (M \lambda, \nu) = \int d{\vec r} \rho_{M,i}( R_{\xi}^{-1}( \theta)  {\vec r}) \, r^{\lambda}Y^{\lambda}_{\nu}( {\hat r})  f^{\pi}( {\vec r}_i )
\nonumber\\  
 &&\,\,\,\,\,\,\, \,\,\,\,\,= \int d{\vec r} \rho_{M,i} ({\vec r}) \, r^{\lambda}Y^{\lambda}_{\nu}( R_{\xi}( \theta) {\hat r})  f^{\pi}( {R_{\xi}( \theta) \vec r}_i )  
 \label{moments}
 \end{eqnarray}
where  
\be
\rho_{M,i}= - e g_i\frac{1}{2} div  (\rho \, r \vec v )
\ee
is the magnetic density, $g_i$ being the gyromagnetic ratio of rotor i, $\rho$ its mass density and $\vec v $ its particle velocity. The electric moments are obtained by changing the magnetic density with the electric density. Only the ${\mathcal R}$-negative rotor contributes to em transition  matrix elements,  because the integral over internal coordinates vanishes when one evaluates the contribution of the ${\mathcal R}$-positive rotor.

{\bf Decay of the ${\mathcal R}-\mbox{negative},1^-_{sc}$ state}. This state can decay only through  electric  transitions, which requires that the $\mathcal {R}$-negative rotor be charged. I therefore assume that rotor 1 has a negative $\mathcal {R}$-symmetry and it  is made of protons. Moreover  it must have negative parity, so that it must be $\sigma$-even. The changes to be made if the neutrons have nonzero effective charges are obvious. To first order in $\theta$, using $f^{-}(\vec{r}_1) $ of Eq.\reff{noncollective} in the moments definition, Eq.\reff{moments}
\be
({\mathcal M}_p^{-})'( E1,\nu) \approx i \theta Q_{p1} <10 | I_{\xi} |1 \nu>
\ee
where  $I_{\xi}$ is the $\xi$-component of the angular momentum in the intrinsic frame and
\be
Q_{p1}=\frac{3}{8} \sqrt{\frac{3}{4 \pi} } \,e_p N_pR \,.
\ee
In the above equation $N_p$  the number of protons and   $R$ the nuclear radius. This result has been obtained by assuming that all the protons contribute. It is known, however, that only protons on the surface do contribute, which is accounted for in the Two-Rotors model by assuming that  because of superfluidity only the part of the rotor external to the maximum  sphere inscribed in the rotor participates in the motion~[\onlinecite{LoIu3}]. In this way one obtains
\be
Q_{p1} \rightarrow Q_{p\delta}=\frac{1}{8} \sqrt{\frac{3}{4 \pi} } \,e_p N_p R \,  \delta 
\ee
where $\delta$ the deformation parameter. There is  a reduction by a factor $ \frac{2}{3} \delta$
which  means that the effective number of protons is $ \frac{2}{3} N_p \delta$.
 Because $ <10 | I_{\xi} |1 0> =0$, the contribution to the electric decay comes entirely from the $|K|=1$ components of the wf so that  
 \be
<\Lambda_{1^-,M,1}|\mathcal {M}(E1,\mu)| \Psi^0_{00}> = i \frac{1}{6}Q_{p\delta} \theta_0 \delta_{M \mu}
\ee
which yields the strength
\be
B(E1; 1_{sc}^- \rightarrow 0^+) =\frac{1}{12} ( Q_{p\delta} \theta_0)^2 \,.
\ee
In a similar way one can evaluate  the strength of decay to the $2^+$ member of the ground state rotational band
 \be
B(E1; 1_{sc}^- \rightarrow 2^+) =\frac{1}{24} ( Q_{p\delta} \theta_0)^2 
\ee
which also comes entirely from the $|K|=1$ components of the scissors mode.

{\bf Decay of the  ${\mathcal R-}\mbox{negative} 1^+_{sc,i}$\, states}. There can be  two scissors states of positive parity, depending on which rotor is  ${\mathcal R}$-negative. In order to have positive parity the  ${\mathcal R}$-negative rotor must be negative also under $\sigma$ inversion.  Using $f^{+}(\vec{r}_1) $ of Eq.\reff{noncollective} in the moments definition, Eq.\reff{moments},
one finds the  components of the  intrinsic magnetic dipole moment for $\mathcal{R}$-negative transitions
\begin{eqnarray}
({{\mathcal M}_i^{+}})' (M1, \pm )&=&    \mp \frac{2}{\pi} { \mathcal C}_i S_3
\nonumber\\
({{\mathcal M}_i^{+}})'(M1, 0) &=&     \frac{2 \sqrt2}{\pi}    { \mathcal C}_i    S_1           
\end{eqnarray}
where
\be
{ \mathcal C}_i= \frac{1}{2} \sqrt {\frac{3}{8\pi} }  \, \frac{eg_i}{2m} 
\ee
 $g_i$ being the gyromagnetic ratio of rotor i.  Notice that the components  $S_1, S_3$  are cartesian, while the components  $({{\mathcal M}_i^{+}})' (M1, \pm ), ({{\mathcal M}_i^{+}})'(M1,0) $ are spherical. In their evaluation I used the approximations $Y^{\lambda}_{\nu}( R_{\xi}( \theta) {\hat r}) \approx Y^{\lambda}_{\nu}({\hat r}), 
 f^{\pi}( R_{\xi}( \theta) {\vec r} ) \approx  f^{\pi}(  {\vec r} ) $.

Let me   first  neglect the term ${\mathcal H}_1$ of the intrinsic Hamiltonian. Consider the decay to the gs. The operator $ S_3$ annihilates a state of zero angular momentum. It is then easy to see that only the $K=0$ component of the intrinsic magnetic moment   contributes  with strength 
\begin{eqnarray}
 B(M1; 1^+_{sc,i} \rightarrow  0^+)= \frac{1}{16 \, \pi^3} \frac{1}{\theta_0^2} \left(\frac{eg_i} {2m} \right)^2 \,. 
\end{eqnarray}
In a similar way one fins that also the decay to the $2^+$ gets contribution only from the $K=0$ component of the scissors mode with branching ratio
\begin{eqnarray}
\frac{B( {\mathcal R}= -1, 1^+_{sc,i}  \rightarrow 2^+ )}{B({\mathcal R}= -1, 1^+_{sc,i}  \rightarrow 0^+  )} =\Big(\frac{1}{2} \frac{\sqrt5 \, C^{10}_{2010}}{ C^{10}_{0010}} \Big)^2 = \frac{1}{2} \,.
\end{eqnarray}
It differs from that of the Alaga [\onlinecite{Alag}] rule  by the factor $1/2$ in the  second member of the equality, and this factor comes from the normalization of ${\mathcal F^2_{M0}} $ in Eq.\reff{normal}, namely from $K$ mixing in the wf of the $2^+$(see the Supplemental Material). 
In the same way, but after inclusion of the term ${\mathcal H}_1$  one gets to order $\alpha$ the strength of  the transition to the $2^+$(in which there is also a $|K|=1$ contribution)
\be
 B(M1; 1^+_{sc,i}   \rightarrow  2^+)\approx\frac{1}{2} ( 1- r_i \alpha )B(M1; 1^+_{sc,i}  \rightarrow  0^+) \label{ratios}
\ee
where
\be
\alpha =\frac{5}{2} \sqrt \pi  \, \frac{{\mathcal J}_2 - {\mathcal J}_1}{ {\mathcal J}_1 + {\mathcal J}_2} \, \theta_0 \,.  \label{Alaga}
\ee
I evaluated  (see the Supplemental material)  also  the electric quadrupole decay strength 
\be
  B(E2; 1^+_{sc,i} \rightarrow 2^+)= \frac{1}{4} \theta_0^2 Q_{i2}^2
\ee
where $ Q_{i2}= - \frac{1}{\pi}\sqrt\frac{3}{10\pi}e_i N_iR^2$, $e_i$  being the effective electric charges and 
$ N_i$ the number of particles of rotor i. It also comes entirely from the $K=0$ component.  It  can be compared with the corresponding strength $B(E2; 1^+_{sc,i}, \mathcal{R}=1 \rightarrow 2^+) $ of the  decay of the $\mathcal {R}$-positive scissors mode[\onlinecite{LoIu}].

Finally  I report by comparison  the magnetic  decay strengths  of the $ \mathcal {R}$-positive scissors mode. The  intrinsic magnetic dipole moment  components in this case are
\begin{eqnarray}
\mathcal{M}'(M1, \mathcal {R}= + 1, \pm) &=& \mp  ( {\mathcal C}_1-  {\mathcal C}_2 ) \, (S_1 \pm S_2)    
\nonumber\\
\mathcal{M}'(M1, \mathcal {R}= + 1, 0) &=& \sqrt2 \, ( {\mathcal C}_1-  {\mathcal C}_2 )\,  S_3         \label{M_+}                
\end{eqnarray}
because both rotors contribute.
The branching ratio of the transition strengths to the  $2^+$ and the ground state, neglecting ${\mathcal H}_{1}$  is
(see the Supplemental Material) is
 \begin{eqnarray}
B(M1; 1^+,  \mathcal {R} = + 1  \rightarrow  2^+)  &=&  \frac{1}{2} B(M1; 1^+,  \mathcal {R}= + 1  \rightarrow  0^+)
  \nonumber\\
  &=&\frac{1}{16 \,  \pi} \frac{1}{\theta_0^2} \left(\frac{e (g_1 - g_2 )   } {2m} \right)^2 . \label{positive}
 \end{eqnarray} 
 The first strength has been evaluated in the present work. A large amount of experimental data is available [\onlinecite{Wess},\onlinecite{Pitz},\onlinecite{Beck2}], which  are in reasonable agreement with the above result. It   comes about in a quite nontrivial way. Both  regions $I$ and $II$ defined before
Eq.\reff{J} give the same contribution, due to the $K=\pm 1$ components of the intrinsic moment: The terms coming  from the operators $S_2, S_3$ in the intrinsic moments, Eqs. \reff{M_+}, are negligible ($O(\theta_0)) $ or cancel out. Moreover both rotors contribute, each one with  with its gyromagnetic ratio, while in the decay of the $\mathcal{R}$-negative scissors mode only the $\mathcal{R}$-negative rotor contributes.

{\bf  Conclusion}. In a previous work [\onlinecite{Palu}] I determined negative $\mathcal {R}$ and  space parity scissors states of the TRM.  Such states were obtained by assuming that each rotor might exist in a positive as well as negative $\mathcal {R}$ parity state. In spite of their two-sided structure, the rotors were assumed to be fully classical objects.  In the present work I treat the rotors as quantal objects as regard of their internal noncollective position variables, allowing also negative parity wr to reflections in a plane through  their symmetry axes. In this way one can construct states of negative $\mathcal {R}$ symmetry with both negative and positive space parity.

Now a few comments concerning the findings of Ref.[\onlinecite{Beck}]. Assuming $\mathcal {R}$ invariance for each rotor

i) The  $1_{sc}^-$ state I constructed might be related to the $1^-$ state reported in Ref.[\onlinecite{Beck}] only if the latter  has  $|K|=0,1$ components of comparable amplitude

ii) The  $1_{sc, i}^+$ of the doublet  I constructed (the proton or neutron rotor being $\mathcal {R}$-negative) have  $|K|=0,1$ components of equal amplitude, in qualitative agreement with the doublet of $1^+$ states of Ref.[\onlinecite{Beck}].These latter states have energies E= 3.159 and  3.173 kev and corresponding branching ratios $B_{1 \rightarrow 2}/ B_{1 \rightarrow 0}= (1+0.36)/2,  (1-0.24)/2$:  The corrections are of opposite sign in the 2 states in agreement with Eq.\reff{ratios}.
This is not trivial: It is a consequence of the structure of the $1^+_{sc,i}$ states, Eq.\reff{parity}. In order to have a quantitative  agreement with the observed values of the  B(M1) strengths, however, the parameter $\alpha $ of 
Eq.\reff{Alaga} should be about 0.3, while with current values of $\theta_0$, $\alpha < 0.1$. 

iii) If instead one requires $\mathcal {R}$ invariance  only for the whole two rotors system, any superposition of the 
states \reff{lambda} is possible in the TRM, because according to Eq.\reff{total} each of them is  odd under the total ${\mathcal R}$ operation.
 
 As already noticed there are several $J=1$ states with energies in the range of the scissors mode and evidence of $K$ mixing [\onlinecite{Zilg,Beck1}]. Perhaps to fully understand the ${\mathcal R}$ nature of such states it might be necessary to measure the em form factors, which should be sensitive to the phase factor $f^{\pi}$ of Eq. \reff{noncollective}.

%A final comment is in order. The Hamiltonian of the TRM has been derived in different ways from many-body Hamiltonians. Dieperink showed [\onlinecite{Diep}] that the Interacting Boson Model in the coherent state approximation for small vibrations of the rotor axes about the $\zeta$ axis reproduces the TRM Hamiltonian in region I. I think that the small vibrations about the $\eta$ axis should reproduce the part of region II, but in the absence of such a result the entanglement of scissors states is not confirmed in the IBM.  In a study of rotational states the TRM Hamiltonian has also been derived [\onlinecite{Bent}] in its classical form \reff{classic}, but  entanglement depends on its quantization which has not been discussed by the Authors. Therefore as far as I know at the moment entanglement  [\onlinecite{Palu2}] is a prediction of the TRM only,and it is conceivable that in actual atomic nuclei proton-neutron roto-oscillations take place only about one axis, the $\zeta$-axis, say. This would not change the predictions for the decay to the ground and the $2^+$ state for the $\mathcal {R}$-positive scissors mode, whose strengths get equal contribution from  regions I and II (of course one should normalize to 1 the wf $\phi_K$). But it would radically change the predictions for the $\mathcal {R}$-negative ones, for which then only the electric or magnetic states would exist.

\section*{Acknowledgements}
I thank  T. Beck and N. Pietralla for a most fruitful correspondence and for communicating some of their results before the publication.


\begin{thebibliography}{11}


\bibitem{LoIu}
N. Lo Iudice and F. Palumbo, Phys. Rev. Lett. {\bf 41}, 1532 (1978);
R. Hilton (Int. Conf. on Nuclear Structure, (Dubna, June 1976) unpublished) also considered the type of collective motion described by the TRM, but he did not either determine its quantum numbers or estimate excitation energy and transition strengths ; T. Suzuki and D. J. Rowe found a $1^+$ state in a RPA calculation that they interpreted as the scissors mode of the TRM, see Sec. IV of Nucl. Phys. A289 ( 1977) 461

\bibitem{Bohle}
D. Bohle, A. Richter, W. Steffen, A. E. L. Dieperink, N. Lo Iudice, F. Palumbo
and O. Scholten, {\it Phys. Lett. B} {\bf 137}, 27 (1984)

\bibitem{LoIu1}
 N. Lo Iudice, La Rivista del Nuovo Cimento,  23, 9 (2000); K. Heyde, P. von Neumann-Cosel and A. Richter, Rev. Mod. Phys. {\bf 82} (2010) 2365

\bibitem{Beck}
T. Beck, J. Beller, N. Pietralla, M. Bhike, J. Birkhan, V. Derya, U. Gayer, A. Hennig, J. Isaak, B. Loher, V. Yu. Ponomarev, A. Richter, C. Romig, D. Savran, M.Scheck, W. Tornow, V. Werner, A. Zilges and M. Zweidinger, Phys. Rev. Lett. {\bf 118} (2017) 212502

\bibitem{Guer}
D. Gu\' ery-Odelin and S. Stringari, {\it Phys. Rev. Lett.} {\bf 83}, 4452 (1999);
E. Lipparini and S. Stringari, {\it ibidem} {\bf 63}, 570 (1989);
A. Minguzzi and M. P. Tosi {\it Phys. Rev. A} {\bf 63}, 023609 (2001);
V.O. Nesterenko, W. Kleinig, F.F. de Souza Cruz and N. Lo Iudice, Phys. Rev, lett. {\bf 83} ( 1999) 57;
P.-G. Reinhard, V.O. Nesterenko, E. Suraud, S. El Gammal and W. Kleinig, Phys. Rev. A {\bf 66} (2002) 013206;
L. Serra, A. puente and E. Lipparini, Phys. Rev. b {\bf 60} (1999) R13966;
K. Hatada, K. Hayakawa and F. Palumbo, Phys. Rev. B {\bf 71}, 092402 (2005);
K. Hatada, K. Hayakawa and F. Palumbo, Eur. Phys. J. B {\bf 77}, 41 (2010) [arXiv:0909.1422], ibid. {\bf 85} (2012) 183; 
E. Lipparini, Atomic Gases, Quantum Dots and Quantum Fluids, World Scientific Publishing Co. Pte. Ltd. (2003)


\bibitem{Mara}
O. M. Marag\' o, S. A. Hopkins, J. Arlt, E. Hodby, G. Hechenblaikner
and C. J. Foot, {\it Phys. Rev. Lett.} {\bf 84}, 2056 (2000)

\bibitem{Ferr}
I. Ferrier-Barbut, M. Wenzel, F. Bottcher, T. Langen, M. Isoard. S. Stringari and T. Pfau, Phys. Rev. Lett. {\bf 120} (2018) 160402

\bibitem{Rocc}
S.M. Roccuzzo, A. Gallemi, A. Recati and S. Stringari, Phys. rev. Lett. 124. 045702

\bibitem{Tanz}
L. Tanzi, J. G. Maloberti, G. Biagioni, A. Fioretti, C. Gabbanini and G. Modugno, Science 371(2021) 1162

\bibitem{Hata}
K. Hatada, K.Hayakawa, C. Marcelli and F. Palumbo, arXiv:1404.4958 (2014), Physical Chemistry Chemical Physics, (2014) DOI:10.1039/C4CP03359K

\bibitem{Brin}
D. M. Brink, Ph. D. thesis, Oxford University, 1955 (unpublished)

\bibitem{Krti}
M. Krticka, F. Becvar, J. Honzatko, I. Tomandl, M. Heil, F. Kappeler, R. Reifarth, F. Voss and K. Wisshak,
Phys. Rev. Lett. 92 (2004) 172501;
A. Schiller, A. Voinov, E. Algin, J.A. Becker, L.A. Bernstein, P.E. Garret, M. Guttormsen, R.O. Nelson, J. Rekstad and S. Siem, Phys. lett. B 633 (2006) 225;
A. Chyzh et al., Phys. Rev. C {\bf 84}  (2011) 014306;
B. Baramsai et al., Phys. Rev. C {\bf 87} (2013) 044609;
T. Renstrom, H. Utsunomiya, H.T. Nyhus, A.C. Larsen, M. Guttormsen, G.M. Tveten, D.M. Filipescu, I. Gheorghe, S. Goriely, S. Hilaire, Y.-W. Lui, J.E. Midtbo, S. Peru, T. Shima, S. Siem and O. Tesileanu, arXiv: 1804.07654 v1 [nucl-ex] 20 Apr 2018

\bibitem{Adek} 
A. S. Adekola et al. Phys. Rev C83 (2011) 34615;
M. Guttormsen, L.A. Bernstein, A. Gorgen, B. Jurado, S. Siem, M. Aiche, Q. Ducasse, F. Giacoppo, F. Gunsing, T. W. Hagen, A.C. Larsen, M. Lebois, B. Leniau, T. Renstrom, S.J. Rose, T.G. Tornyi, G.M. Tveten, M. Wiedeking and J.N. Wilson, Phys. Rev. C {\bf 89} (2014) 014302

\bibitem{Hame}
M. Hamermesh, Group Theory and its Applications to Physical Problems, p. 36, Addison- Wesley Publishing Company, INC

\bibitem{Palu}
F. Palumbo, Nucl. Phys. A, 983 (2019) 64

\bibitem{Zilg}
A. Zilges, P. von Brentano, A. Richter, R. D. Heil, U. Kneissel, H.H. Pitz and C. Wesselborg,  Phys. Rev. C 42(1990) 1945

\bibitem{Beck1}
T. Beck et al., Phys. Rev. Lett. 125 (2020) 092501

\bibitem{Vill}
F. M. H. Villars, Nucl. Phys. {\bf 3} (1957) 240; Ann. Rev. Nucl. Sci. {\bf 7} (1957) 185

\bibitem{Sche}
W. Scheid and W. Greiner, Annals Phys. 48 (1968) 493

\bibitem{DeFr}
G. De Franceschi, F.Palumbo and N. LoIudice, Phys. Rev. C{\bf 29} (1984) 1496

\bibitem{Alag}
G. Alaga, K. Alder. A. Bohr and B. Mottelson, Dan. Mat. Fys. Medd. 29 (1955) 

\bibitem{LoIu3}
N: Lo Iudice, F. Palumbo, A. Richter and H. J. Wortche, Phys. Rev. C42 (1990) 241

\bibitem{Wess}
C. Wesselborg, P. Von Brentano, K. O. Zell, R. D. Heil, H. H. Pitz, U. E. P. Berg, U. Kneissl, S. Lindenstruth, U. Seemann, and R. Stock, Phys. Lett. B (1988) 22

\bibitem{Pitz}
H. H. Pitz, R. D. Heil, U. Kneissl, S. Lindenstruth, U. Seemann, R. Stock, C. Wesselborg, A. Zilges and P. Von Brentano, S.D. Hoblit and A. M. Nathan, Nucl. Phys. A509 (1990) 587

\bibitem{Beck2}
K.E.Ide et al. Phys. Rev. 2021


%\bibitem{Diep}
%A. E. L. Dieperink, Progtr. Part. Nucl. Phys. 9, (1983) 121

%\bibitem{Bent}
%W. Bentz, A. Arima, J. Enders, A. Richter and J. Wambach, Phys. rev. C 84 (2011) 014327

%\bibitem{Palu2}
%F. Palumbo, Phys. Rev. C 93 (2016) 034331



%\end{references}
\end{thebibliography}
\end{document}